\documentstyle[sprocl]{article}

\def\be{\begin{equation}}
\def\ee{\end{equation}}
\def\bea{\begin{eqnarray}}
\def\eea{\end{eqnarray}}
\begin{document}

\title{COSMOLOGICAL PARAMETERS AND THE CASE FOR COLD DARK MATTER}

\author{MICHAEL ROWAN-ROBINSON}

\address{Astrophysics group, Blackett Laboratory, Imperial College,\\ Prince Consort Rd, London SW7 2BZ\\E-mail: mrr@ic.ac.uk}


\maketitle\abstracts{ Determinations of the main cosmological parameters are reviewed and
the implications for cold dark matter discussed.  There is no longer an age problem for
an $\Omega_o$ = 1, $\Lambda$ = 0 model and , if anything, there is now an age problem for 
low $\Omega_o$, $\Lambda > 0$ models.  Large scale structure and CMB fluctuation data are
best fitted by a mixed dark matter $\Omega_o$ = 1 universe.  Difficulties for this model
with cluster evolution, the baryon content of clusters, high z Lyman $\alpha$ 
galaxies, and the evidence from Type Ia supernovae favouring low $\Omega_o$, 
$\Lambda > 0$ models, are discussed critically.}

\section{Introduction}
Most dark matter searches rely on a substantial fraction of the dark halo of our Galaxy being
made up of cold dark matter, and that this cold dark matter consists of the neutralino. Although 
there is a logical possibility that all of the dark halo is baryonic (white dwarf stars, say),
there is no viable galaxy formation scenario in which this could be the case.  Moreover it has 
proved extremely difficult to construct an astrophysical scenario in which even as much as 20 $\%$
of the halo is in the form of white dwarf stars (the possibility that such a large fraction could
be low mass hydrogen-burning stars, brown dwarfs or Jupiters is already ruled out).  Currently it
seems more plausible to assume that the microlensing events seen towards the Magellanic Clouds
are in fact due to lensing by stars in the Clouds.

In this paper I review the current evidence on cosmological parameters and assess their relevance to
cold dark matter searches.  Section 2 reviews the main cosmological parameters $H_o, T_o, \Omega_b,
\Omega_o, \Lambda$.  Section 3 briefly discusses evidence from large-scale structure and other arguments
on the nature of the dark matter. 

\section{Cosmological parameters}

\subsection{Hubble constant, $H_{o}$}
The distance scale and the Hubble constant ware reviewed by Rowan-Robinson (1985, 1988), and 
the Hubble constant and other cosmological parameters were 
reviewed by Rowan-Robinson (1997).  There appears to be a growing concensus that the Hubble 
constant lies in the range 55-75 km$/$s$/$Mpc.  It is becoming clear that only primary distance
methods, which are geometrical methods, are calibrated within our Galaxy or have a theoretical 
underpinning, really add to our knowledge of the Hubble constant.  Here I summarize recent 
developments since 1996.

{\bf Cepheids:}
Freedman et al (1998) have summarized the results of the HST Key Program using
Cepheids in galaxies out to 20 Mpc to determine the Hubble constant and found
$H_o$ = 75 $\pm$15 km$/$s$/$Mpc.  The local flow is still a major uncertainty 
for this method.

{\bf Supernovae Type Ia:}
Branch (1998) has given an excellent review of the current situation.
Sandage et al (1994) concluded from HST observations of Cepheids in two galaxies
in which Type Ia supernovae had occurred that $H_o$ = 55 $\pm$8 km$/$s$/$Mpc.  Riess
et al (1995) used the relation between luminosity at maximum light and decay rate
over the next 15 days found by Phillips (1993) that $H_o$ = 67 $\pm$7 km$/$s$/$Mpc.
Tammann and Sandage (1995) disputed the Riess et al analysis and argued that
the slope of the claimed relation could not be nearly as steep as claimed.  Hamuy
et al (1996) analyzed a large sample of nearby supernovae and gave several possible
solutions for the luminosity-decay rate relation.  However their preferred solution 
was much less steep than the Phillips/Riess et al value, and consistent with the
Tammann and Sandage limit.  Their revised value for the Hubble constant was
$H_o$ = 63 $\pm$4.5 km$/$s$/$Mpc.  Saha et al (1997) analyzed 7 Type Ia supernovae
with HST Cepheid distances and found $H_o$ = 58 $\pm$8 km$/$s$/$Mpc (61 if they
included a relationship between $M_B$ and $\Delta M_{15}$).  Branch et al (1996)
found $H_o$ = 57 $\pm$4 km$/$s$/$Mpc from a colour matched sample of supernovae and
Tripp (1997) found $H_o$ = 60 $\pm$5 km$/$s$/$Mpc from a sample matched according to
$\Delta m_{15}$.  Freedman et al (1997) found $H_o$ = 67 $\pm$8 km$/$s$/$Mpc for
supernovae in the Fornax cluster.  Riess et al (1996) used a template method ('MLCS')
to obtain $H_o$ = 64 $\pm$6 km$/$s$/$Mpc.  

Several groups have brought theoretical models to bear on the determination of
the Hubble constant using Type Ia supernovae.  Hoflich and Khokhlov (1996) compared 
26 supernovae with model light curves and found $H_o$ = 67 $\pm$9 km$/$s$/$Mpc.  Branch (1998)
suggests this should be revised to 56 $\pm$ 5.  The same two authors found $H_o$ = 55
if they included a theoretical version of the  $M_B$ - $\Delta M_{15}$ relation.  
Nugent et al (1995) fitted non-LTE model spectra to observations and found $H_o$ = 60 + 14,-11.
  In a detailed review of these and other 
determinations, Branch (1998) concluded that the best estimate for the Hubble constant
from Type Ia supernovae was $H_o$ = 60 $\pm$5 km$/$s$/$Mpc. 

{\bf Supernovae Type II:}
Schmidt et al (1994) found $H_o$ = 73 $\pm$13 km$/$s$/$Mpc using Type II supernovae.

{\bf Gravitational lens time delay:}

A recent analysis by Falco et al (1997) of the gravitational lens time delay system
0957+561 and found $H_o$ = 62 $\pm$7 km$/$s$/$Mpc.  There is a need for other systems
of this type to be found and studied to test for systematic effects.

{\bf Sunyaev-Zeldovich effect:}
Birkinshaw et al (1994) gave $H_o$ = 55 $\pm$17 km$/$s$/$Mpc for 2 clusters, Myers et al (1997)
gave results for a further four clusters, which average to 54 $\pm$ 14, 
and Hughes and Birkinshaw (1998) give a value for CL00016+16 of 47 (+23,-15).

\medskip
To summarise recent estimates of $H_{o}$ by primary methods:
\begin{itemize}
\item $H_o$ = 75 $\pm$ 15,  Cepheids in Virgo, Fornax (Freedman et al 1998)

\item $H_o$ = 60 $\pm$ 5,  Cepheids + SN Ia (Branch 1998)

\item $H_o$ = 73 $\pm$ 13, SN II, Baade-Wesselink (Schmidt et al 1994)

\item $H_o$ = 62 $\pm$ 7, gravitational lens time delay (Falco et al 1997
)						
\item $H_o$ = 54 $\pm$ 14, Sunyaev-Zeldovich effect (Myers et al 1997).

\item straight average of these 5: {\bf $H_o$ = 65 $\pm$ 8  km$/$s$/$Mpc}

\item weighted average of these 5: {\bf $H_o$ = 62 $\pm$ 4  km$/$s$/$Mpc}

\end{itemize}

If all clusters and groups with reliable distances are combined in a 
manner similar to CDL, RR88, a value of 65 is found.  If only galaxies 
with distances greater than 100 Mpc are used, a value of   
$H_o$  = 62 $\pm$ 5 km$/$s$/$Mpc is found.  I adopt  {\bf $H_o$ = 65 $\pm$ 8  km$/$s$/$Mpc}
as a conservative estimate which encloses all proposed values within the 2-$\sigma$ range.

\subsection{Age of the universe, $t_{o}$}
The turnoff point of globular clusters has in the past generally yielded ages for the oldest
globular clusters of 14-18 Gyr.  Recently Jimenez et al (1996) have proposed a
new method of determining globular cluster ages, based on a comparison of the
horizontal branch morphology with the distribution of stars near the tip of the
red giant branch.  They have concluded that the age of the oldest clusters
is 13.5 $\pm$ 2 Gyr.  

The HIPPARCOS data has a considerable impact on both the Hubble constant
and on the age of the universe (Feast and Catchpole 1997, Feast and Whitelock 1998,
Sandage and Tammann 1998, Gratton et al 1998).
There are 220 Cepheids with parallaxes from Hipparcos.  These have the
consequence that the
Cepheid (Population I) distance scale is increased by 8-10$\%$.

This in turn has an immediate impact on distances derived from Cepheids, eg HST 
Key Program estimate of $H_o$ would be reduced by 10$\%$
(but see Madore and Freedman 1997).
The revised distances to LMC and M31 increase the derived mean luminosity of 
RR Lyrae stars, hence increasing the Population II distance scale.
The distances of the old metal-poor globular clusters have therefore been 
underestimated by 8-10 $\%$.
Finally this means their ages have therefore been overestimated and should be 
reduced to 11-12 
Gyr (turnoff point corresponds to higher luminosities, therefore to more 
massive and younger stars).  In a very extensive series of Monte Carlo simulations,
Chaboyer et al (1998) conclude that {\bf $t_{o}$ = 11.5 $\pm$1.3 Gyr}.  To this must be 
added the time since the Big Bang for the formation of globular clusters, likely to be 
in the range 0.1-1.5 Gyr (for formation at z = 20 - 3 in an $\Omega_o$ = 1 universe),
but could be more if formation was at even lower redshift, or in a low $\Omega_o$
universe.

\subsection{Baryonic density,  $ \Omega_{b}$}
Big Bang nucleosynthesis of the light elements D, $^{3}He, ^{4}He, ^{7}Li$, gives a reasonably
accurate estimate for $ \Omega_{b}$ = 0.03 $\pm$ 0.006 $(65/H_{o})^{2}$ (Walker et al 1991).

This result has been under pressure in two directions in the past year or so.
Tentative detection of a high D abundance in high resolution quasar absorption line
studies, using the Keck telescope, threatened to push $ \Omega_{b}$ to much lower values.
Recently Tytler et al (1997) have shown that this high D abundance system is much more
likely to be due to a second hydrogen absorption.  At the same time Tytler et al have found 
evidence in other systems for a lower D abundance, leading to an estimate for

$ \Omega_{b}$ = 0.05 $\pm$ 0.01 $(65/H_{o})^{2}$
  
\subsection{Total density,  $ \Omega_{o}$}

The most important measurements of the total density have come from all sky 
IRAS galaxy redshift surveys, specifically the 1.94 Jy survey (Strauss et al 
1990), the QDOT survey (Rowan-Robinson et al 1990), and the 1.2 Jy survey (Fisher et al 1992).
The QDOT team have
 now determined redshifts for the entire IRAS PSC to 0.6 Jy, the PSCz survey (Saunders et al 1996). 
Table 1 summarizes the main IRAS galaxy redshift surveys made to date. 

The IRAS galaxy redshift surveys have been used for a wide variety of cosmological 
studies. Excellent reviews have been given by Dekel (1994) and Strauss and Willick 
(1995).  Table 2 summarizes some of the work in the different scientific
areas carried out with the different surveys.  For many of these statistical
methods, the key quantity that is determined is the quantity $$\beta = \Omega_o^{0.6}/b$$ 
where $\Omega_o$ is the cosmological density parametere and b is the bias parameter.  
Table 3 summarises the value of $\beta$ determined in each of these studies, and also averages
these values horizontally (by method) and vertically (by survey).  The mean value
for all methods is $$\beta = 0.8$$ with a 1-sigma uncertainty of 0.15. This corresponds
to $\Omega_o$ = 0.7 $\pm$ 0.2 for b = 1, or b = 1.25, +0.29, -0.20, for $\Omega_o$ = 1.
The most recent estimate from a comparison of the IRAS 1.2 Jy sample with the POTENT
density field derived from the Mark III catalogue of peculiar velocities, gave $\beta = 0.89 \pm 0.12$
(Sigad et al 1997), again consistent with $\Omega_{o}$ = 1.

It is interesting to note that there is no gross inconsistency in any of the horizontal 
means or of the vertical means, suggesting that there is no gross systematic
error in any one of the methods or in any one of the surveys.  For example Lauer and 
Postman (1994), Coles and Ellis (1994) and Plionis et al (1995) have
suggested that there are problems with non-convergence of the IRAS dipole, and 
Hamilton (1995) has suggested that there is a problem with the QDOT surveys.  These
claims are not borne out by the consistency of Table 4. 

The problem that IRAS surveys undersample elliptical galaxies, and therefore the dense 
cores of clusters, has been addressed by Strauss et al (1992).  Correction
for the undersampling leads to only very small changes to the derived values of $\beta$, 
because the cores of rich clusters comprise only a small fraction of the
mass of superclusters. 

In conclusion, many different analyses of independent IRAS galaxy redshift surveys 
suggest that IRAS galaxies provide a relatively unbiassed sample of the matter
in the universe and that a universe with $\Omega_o$ = 1 is favoured.  A universe 
with $\Omega_o <$ 0.3 appears unlikely.  This appears to be in conflict with X-ray estimates
of the baryon content of clusters (White et al 1993),
which yields $\Omega_{o} \leq$ 0.3.  A study of the abundance of galaxy clusters and
its evolution yields $\Omega_{o} = 0.45 \pm 0.2$ (Eke et al 1998).    

Larger redshift surveys, like the PSC-z survey, may help to resolve some of these problems.
At the moment I conclude that {\bf $\Omega_o$ = 0.7 $\pm$ 0.2}.

\begin{table}
\caption{IRAS galaxy redshift surveys to date.} \label{tbl-1}
\begin{center}\scriptsize
\begin{tabular}{crrrrrrrrrrr}
&&&\\
name & definition & N & reference \\
\hline
 & & & \\
QDOT & whole sky, & 2187 & Rowan-Robinson et al 1990 \\
 & 1 in 6 to 0.6 Jy & & \\
 & & & \\
2 Jy & whole sky, 1.94 Jy & 2685 & Strauss et al 1990 \\
 & & & \\
1.2 Jy & whole sky, 1.2 Jy & 5339 & Fisher et al 1992 \\
 & & & \\
PSC-z & whole sky, 0.6 Jy & 15,000 & Saunders et al 1996 \\
 & & & \\
 & & & \\
\end{tabular}
\end{center}

\end{table}

\begin{table}
\caption{Scientific results from IRAS galaxy redshift surveys.} \label{tbl-1}
\begin{center}\scriptsize
\begin{tabular}{crrrrrrrrrrr}
method & NGW & BGS & QDOT & 2 Jy & 1.2 Jy & FSS \\
\hline
 & & & & & & \\
dipole&Yahil& &RR et al&Strauss &Strauss & \\
 &et al 86& &90&\& Davis 88&et al 92& \\
 & & & & & & \\
vel. field& & &Kaiser&Dekel&Fisher& \\
vs. dens.& & &et al 91&et al 93&et al 94& \\
fields& & &Taylor \&&Roth 94&Nusser \&& \\
 & & & RR 94& &Davis 94& \\
 & & & & &Willick & \\
 & & & & &et al 96& \\
 & & & & & Sigad & \\
 & & & & & et al 97 &\\
 & & & & & & \\
z-space& & &Cole et al 95&Hamilt 93&Fisher& \\
distn& & &Hamilt 95&Fry \& Gaz 93&et al 94a& \\
$\sigma-\pi$ & & & & & & \\
 & & & & & & \\
power& & &Taylor&Fisher&Peacock& \\
spectrum& & &\& RR 92&et al 93&\& Dodds 94& \\
P(k)& & & & &Feldman& \\
 & & & & &et al 94& \\
 & & & & &Cole et al 94& \\
 & & & & & & \\
spherical& & & & &Fisher et al 94b& \\
harmonics& & & & &Heavens\& & \\
 & & & & &Taylor 95& \\
\end{tabular}
\end{center}
\end{table}

\begin{table}
\caption{IRAS estimates of $\beta = \Omega_o^{0.6}/b$.} \label{tbl-1}
\begin{center}\scriptsize
\begin{tabular}{crrrrrrrrrrr}
method & QDOT & 2 Jy & 1.2 Jy & $<\beta>$ \\
\hline
dipole & 0.94 $\pm$ 0.2 & & 0.55 +0.20-0.12 & 0.75 \\
 & R90,L95 & & S92 & \\
 & & & & \\
vel. field & 0.86 $\pm$ 0.14(K91) & 1.28 $\pm$ 0.34(D93) & 0.6 (ND94) & 0.80 \\
vs. dens field & 0.83 $\pm$ 0.10(T94) & 0.6 (R94) & 0.55 $\pm$ 0.13(W95) & \\
 & & & 0.89 $\pm$0.12(S97) & \\
 & & & & \\
z-space distn & 0.54 $\pm$ 0.3(C95) & 0.69 $\pm$ 0.27(H93) & 0.45 $\pm$ 0.22(F94a) & 0.60 \\
 & & 0.84 $\pm$ 0.45(FG93) & 0.52 $\pm$ 0.3(C95) & \\
 & & & & \\
power spectrum & & & 1.0 $\pm$ 0.2(PD94) & 1.0 \\
 & & & & \\
spherical & & & 0.94 $\pm$ 0.17(F94b) & 1.0 \\
harmonics & & & 1.1 $\pm$ 0.3(HT94) & \\
 & & & & \\
 & & & & \\
 & & & & \\
$<\beta>$ & 0.80 & 0.85 & 0.73 & {\bf 0.80 $\pm$ 0.15} \\
\end{tabular}
\end{center}
\end{table}

\subsection{Cosmological constant, $\Lambda$}
There has been a great deal of excitement about the possibility that Type Ia supernovae
yield positive evidence that $\Lambda >$ 0 (eg Krauss 1998). The observational situation 
on $\Lambda$ was reviewed recently by Fort and Mellier (1998).  
The statistics of gravitationally lensed quasars provides a strong
constraint on the value of $\lambda = \Lambda/3H_{o}^{2}$ (Turner et al
1994, Turner 1990, Kochanek 1996).  Kochanek gives a firm 2-$\sigma$ limit of 
$\lambda <$ 0.65.

Other methods which have been applied include statistics of quasar pair separations 
(Myungshin et al 1997), geometry of gravitational arcs and arclets (reviewed
by Fort and Mellier 1998), and statistics of quasar absorption line systems
(Turner and Ikeuchi 1992).

Recently two groups have published results on the Hubble diagram for Type Ia
supernovae, with over 100 supernovae now discovered at z $>$ 0.3 
(Schmidt et al 1998, Garnavich et al 1998, Riess et al 1998, Perlmutter et al 1998).  Both groups claim that
models with positive cosmological constant are preferred, and that models with
$\lambda = 0.7, \Omega_{o} = 0.3$ provide the best fit to the data.  This seems 
marginally inconsistent with the limits from statistics of 
gravitational lensed quasars.  The strength of the signal is that Type Ia supernovae at redshift
0.3-0.9 are about 0.25 magnitudes fainter than local supernoave, if an $\Omega_o$ = 1 
Einstein-de Sitter universe is assumed.  Claims that this is a 7-8 $\sigma$ effect 
therefore depend on a very precise homogeneity of Type Ia supernovae.  Looking back to, for example,
the review by Branch and Miller (1993), in which the rms scatter of the absolute magnitude of
Type Ia supernovae at maximum light was given as $\sigma$ = 0.36 after judicious omission of 
anomalous objects, this does seem a remarkable claim.  The key element
in reducing the scatter in Type Ia supernova absolute magnitudes at maximum light has been the
correlation between absolute magnitude and decline rate ($M_B - \Delta m_{15}$), discussed
above.  If one looks at the paper by Hamuy et al (1996) where this relation is
established for 29 local supernovae, one finds that the situation is not quite
as impressive as has been presented.  It would seem reasonable that to talk about a relationship between
the absolute magnitude at maximum and the decline rate over the next 15 days, it
would be necessary to have detected the calibrating supernovae prior to maximum.
In fact only 10 of the local supernovae were first observed at least one day
before maximum.  For these 10 there is indeed a  $M_B - \Delta m_{15}$ relation, but its
significance is much reduced.  If we derive the calibration from these 10 local
supernovae and apply it to the distant supernovae, the significance of the signal
is reduced from the claimed 7-8 $\sigma$ to only 2-3 $\sigma$ (depending on how many
distant supernovae are used).  It appears that the calibrating relation needs to be placed 
on a much stronger basis with nearby supernovae before it can be used to establish
the reality of a cosmological constant.  A good test of homogeneity would be
to find several supernova in a high redshift cluster.

There are also some theoretical uncertainties.  Since we do not know for certain whether
nearby supernovae are due to white dwarf deflagration or to white dwarf mergers, there
is the possibility that the proportion of these two types changes with epoch and this could affect
the mean absolute magnitude.
Hoflich et al (1998) also point out that uncertainties and
evolution of the initial composition in supernovae can have a major effect on the
determination of cosmological parameters using supernovae.

\subsection{Summary on cosmological parameters}
My best estimates of the cosmological parameters discussed above were:

	$H_{o} = 65 \pm 8$ km$/$s$/$Mpc
		corresponding to Hubble time  $\tau_{H}$ = 15.1 Gyr

	$t_{o} = 11.5 \pm$ 1.3 Gyr (plus 0.1-1.5 Gyr for globular cluster formation)

	$\Omega_{o} = 0.7 \pm$ 0.2

with  $\lambda$ undetermined, but probably $\leq$ 0.7.

Three scenarios consistent with these values are:
\begin{itemize}
\item (1) $\Omega_o = 1, \lambda = 0, H_o = 60 km/s/Mpc, t_o = 11$ Gyr.
\item (2) $\Omega_o = 0.3, \lambda = 0.7, H_o = 70 km/s/Mpc, t_o = 13.5$ Gyr,
\item (3) $\Omega_o = 0.3, \lambda = 0, H_o = 65 km/s/Mpc, t_o = 12.3$ Gyr.
\end{itemize}

The first and third are in conflict with the Type Ia supernova estimates of $\lambda$.  The second and third
are only marginally consistent with estimates of $\Omega_o$ form large-scale flows. The first and second are
uncomfortable with a Hubble constant of 65 $km/s/Mpc$, implying, respectively, too low and too high ages 
for the universe.    Can other arguments settle the issue ?

\section{Large-scale structure and other arguments}

Gawiser and Silk (1998) have examined a range of cosmological scenarios, most of which are based on 
cold dark matter, and compared them with the available data on the power spectrum of density fluctuations
derived both from CMB anisotropies and from clustering of galaxies.  They reach the same conclusion as
an earlier study of this type by Taylor and Rowan-Robinson (1992), that the only scenario consistent with all
the available data is an $\Omega_o$ = 1 mixed dark matter model (the model they selected has 20$\%$ hot
dark matter.  Caldwell (1998) has reviewed all the available data on neutrino masses and concludes that 
only a 4-neutrino model, with two pairs of degenerate neutrinos and including a low-mass sterile neutrino,
can fit all the existing measurements and limits.  This model would be consistent with the required $\Omega_{\nu}$.
The second of the three possible scenarios summarized at the end of section 2 is definitely a significantly poorer fit to the data, 
and the third scenario is probably ruled out by the CMB anisotropies, which favour a value for 
$\Omega_o + \lambda \simeq$ 1.

Gawiser and Silk also enumerate the various problems for an $\Omega_o$ = 1 universe: 
\begin{itemize}
\item {\bf cluster evolution:} strong negative evolution would be expected in the space density of rich clusters
in an $\Omega_o$ = 1 universe because growth of structure continues to the present day.  This has been claimed
not to have been seen by Bahcall et al (1997), Fan et al (1997), Carlberg et al (1997), Eke et al (1998).  See
however Burke (1998).

\item {\bf the cluster baryon fraction:} Many groups have confirmed the original finding of White et al (1993)
that the baryon content of clusters is too high for an $\Omega_o$ = 1 universe ( White and Fabian (1995), 
Loewenstein and Mushotzky (1996), Mulchaey et al (1996), Evrard (1997)).

\item {\bf high-redshift Lyman $\alpha$ galaxies:} The mixed dark matter scenario has problems with the 
abundance of high redshift Lyman $\alpha$ galaxies, because it predicts too little power at small scales.

\item {\bf abundance of cluster arcs:} Bartelmann et al (1998) find that $\Omega_o$ = 1 models predict too
few strong lensing arcs in clusters of galaxies.
\end{itemize}

In spite of all these arguments in favour of a low value of $\Omega_o$, Gawiser and Silk still conclude that 
the evidence from large-scale structure outweighs this.  It could be noted that
several of these arguments may be affected by the still uncertain details of
galaxy formation.  It seems premature to conclude that a low value of
$\Omega_o$ and a positive value for $\lambda$ have been established.  

In conclusion there are two viable scenarios consistent with the evidence on cosmological parameters:
an $\Omega_o = 1, \lambda = 0$ universe or a  $\Omega_o = 0.3, \lambda = 0.7$ universe.  Although a number
of lines of evidence, including high redshift Type Ia supernovae, favour the latter, the evidence from
large-scale structure still favours the former.
In either case, it is likely that most of the dark matter in
the halo of our Galaxy is in the form of cold dark matter.

\end{document}